\documentclass[12pt,a4paper]{article}
\usepackage{amsmath}
\usepackage{amsfonts}
\usepackage{cite}
\usepackage{graphicx}
\usepackage{epsfig,multicol,bbm}

\begin{document}
\begin{titlepage}

\vfill
\begin{flushright}
ACT-06-10, MITPA-10-17
\end{flushright}

\vfill

\begin{center}
   \baselineskip=16pt
   {\Large\bf Hitchin Equation, Irregular Singularity, and $N=2$ Asymptotical Free Theories}
   \vskip 2cm
     Dimitri Nanopoulos $^{1,2,3}$, and Dan Xie $^{1}$       \vskip .6cm
      \begin{small}
      $^1$\textit{ George P. and Cynthia W.Mitchell Institute for Fundamental Physics,
Texas A\&M University, College Station, TX 77843.}
        \end{small}\\*[.6cm]
      \begin{small}
      $^2$\textit{Astroparticle physics Group, Houston Advanced Research
Center (HARC), Mitchell Campus, Woodlands, TX 77381, USA.}
        \end{small}\\*[.6cm]
       $^3$\textit{Academy of Athens, Division of Nature Sciences, 28
panepistimiou Avenue, Athens 10679, Greece.}
\end{center}

\vfill
\begin{center}
\textbf{Abstract}\end{center}

\begin{quote}In this paper, we study  irregular singular solution to Hitchin's equation and
use it to describe four dimensional $N=2$ asymptotically free gauge theories. For $SU(2)$ $A$ type quiver,
two kinds of irregular singularities besides one regular singularity are needed for the solution of Hitchin's equation; We then classify irregular singularities needed for the general $SU(N)$ $A$ type quiver.
 \end{quote}

\vfill

\end{titlepage}

\section{Introduction}
Recently, there are a lot of exciting progress about four dimensional $N=2$  field theories.
Initially, Argyres and Seiberg \cite{Argy} found a remarkable duality for $N=2$ $SU(3)$ gauge theory with
six fundamental hypermultiplets in the infinite coupling limit: In the dual description, a weakly
coupled $SU(2)$ gauge group appears and the mysterious strongly coupled $E_6$ Superconformal \cite{min1}
field theory also appears. This kind of duality is later generalized by Gaiotto to a large
class of $N=2$ superconformal field theories \cite{Gaiotto1}. The crucial idea of Gaiotto is to
engineer four dimensional $N=2$ theories from six dimensional $A_N$ $(0,2)$ theory \cite{witten1}
and compactify six dimensional theory on a Riemann surface. To describe conventional $N=2$ gauge theories,
the defects need to be turned on several marked points on Riemann surface, the
four dimensional gauge theories and the Seiberg-Witted curve are determined entirely
by the punctured Riemann surface and the information encoded in the defects. The Argyres-Seiberg
duality is interpreted as the different degeneration limits of this punctured Riemann surface.
Since Riemann surface with punctures is the natural arena for two dimensional conformal field theory, one may
wonder if there is any connection between four dimensional gauge theory and two dimensional
conformal field theory. AGT \cite{Gaiotto2} found a surprising isomorphism between the Nekrasov partition
\cite{part} function of the gauge theory with conformal block of the Liouville theory in the $SU(2)$ case.

Since there is no lagrangian description for the six dimensional theory, it is hard to
get some concrete information from this construction. However, it seems that Hitchin's equation \cite{hitchin1,hitchin2} plays a central role in all of these developments. In paper \cite{dan}, we argue that Hitchin's equation
is the BPS equation when six dimensional theory is compactified on Riemann surface. In the case
of Superconformal field theories (SCFT), the solution of Hitchin's equation have to allow regular singularity
(the fields have only simple pole) at the punctures.
The moduli space $M$ of the solution to Hitchin's equation has a lot of important implications. The moduli space is a hyperkahler manifold and has a lot of complex structures parameterized by a sphere.
In one distinguished complex structure, every point on the moduli space describes a Higgs bundle on Riemann surface;
In this complex structure, the moduli space can be written as Hithcin's fibration which can be identified with
the Seiberg-Witten fibration of four dimensional gauge theory \cite{witten4,witten5}. When four dimensional theory
is compactified on a circle $S$ with radius $R$ down to three dimensions, the moduli space is
the Coulomb branch target space of low energy effective theory of 3d theory \cite{3d}. The hyperkahler metric of the moduli space plays an crucial role in proving the wall crossing formula \cite{moore,moore1}. When
three dimensional theory is compactified on another circle $S^{'}$ to two dimension,
Hitchin's moduli space becomes the target space of the two dimensional sigma model.

Hitchin's system is a complete integrable system. It is known long time ago that Seiberg-Witten solution
is related to integrable system \cite{inte1, inte2, inte3}. It is shown for $N=2$ superconformal
theory with adjoint matter, the Seiberg-Witten solution is related to Hitchin's equation defined
on a torus with one simple singularity\cite{donagi}. In general,
Seiberg-Witten solution is related to integrable systems appearing
in other context of physics; Some of those integrable
systems are also identified with Hitchin's integrable systems \cite{nek1}.
Recently, there is another surprising relation about
gauge theory and the quantization of integrable system associated with Seiberg-Witten solution
\cite{nek2, nek3, nek4}. Basically, if four dimensional $N=2$ theory is compactified
on $\Omega$ background to two dimensions, the effective two dimensional twisted superpotential
is argued to be the same as the Yang-Yang function of the integrable system which is the
key for the quantization of integrable system.

The intriguing relations of $N=2$ gauge theories with the two dimensional conformal field theories and
quantization of integrable system are explained elegantly by Nekrasov and Witten \cite{nek5}
by compactifying four dimensional theory on different $\Omega$ backgrounds to two dimensions and then
use the branes to study quantization of subspace of Hitchin's moduli space.
Hitchin's moduli space plays an important role here: it is the integrable system associated with
four dimensional gauge theory; it is the target space of the two dimensional sigma model, its submanifold
can be identified with the Teichmullar space whose quantization gives Liouville theory.

Hitchin's equation also plays a central role in recent developments of physical approach to
Geometric Langlands Program \cite{GL,witten2, wild, gukov, gukov1}. We think that problem
is closed related to four dimensional $N=2$ gauge theory.
One hint is coming from six dimensional construction. We first compactify six
dimensional theory on a Riemann surface with or without defects insertion to four dimension and then
compactify on a two torus down to two dimensions, the final theory is a two dimensional sigma model
whose target is Hitchin's moduli space. On the other hand, if 6d theory is first compactified on a two torus, we
get a four dimensional $N=4$ theory, we then compactify on Riemann surface with or without punctures,
the final theory is also two dimensional sigma model with Hitchin's moduli space as target.
The punctures are described physically by surface operators in Geometricl Langlands Program.
It seems that the defects in $N=2$ context are just the
surface operators in $N=4$ context, this is confirmed for the tamed ramification case \cite{dan}.

In this work, we will write down the Hitchin's system description for the general asymptotical free
$A$ type quiver, (see early discussion in \cite{kapustin, kapustin1} and more recently
\cite{moore1}of mapping the Seiberg-Witten curve to Hitchin's system). We start directly from Hitchin's equation and show that we must have irregular singularity for the solution to account for all the UV parameters
for the gauge theories.
There are several motivations to find the explicit form for the singularity
to the Hitchin's equation. The
first motivation is to classify the singularity types needed to describe the conventional $A$ type quiver
and use these singularities to construct new theories, this is the first step of classifying
four dimensional $N=2$ theories. The second motivation is to determine the exact information encoded in
the singularity, this information is important to extract what kind of states
can be put on the puncture in two dimensional conformal field theory (see the SCFT case in \cite{primary}).
The third motivation is
to attach an Hitchin's integrable system to every $N=2$ gauge theory, the
quantization of the integrable system can be derived by studying gauge theory with $\Omega$ deformation;
We will show that some of the conventional integrable systems are related to Hitchin's system.
The fourth motivation is to study three dimensional theory, since Hitchin's moduli space
is the Coulomb branch of three dimensional theory, knowing the explicit form of moduli
space will help us understand property of three dimensional theories, for instance, we may
discover new mirror pairs \cite{KS,GW1,GW2,new}, etc.

For the SCFT, only regular singularities are required (for canonical representation). For
the non-conformal case, the irregular singularity is needed. Irregular singularity is studied by Witten \cite{wild}
in the context of wild ramification of Geometric Langlands Program. The form of irregular singularity
needed for $N=2$ gauge theory is not the standard form studied in the context of Geometric Langlands
Program (Our case is discussed briefly by Witten \cite{wild}). Generically, there is cut around the
irregular singular point. In the $SU(2)$ case, we show that for general $SU(2)$ A type quiver, there are three
kinds of singularities for the Higgs field: simple singularity, order two singularity with leading order coefficient semi-simple and order two singularity with leading order coefficient nilpotent. For the $SU(N)$ case, the form of the irregular singularity is determined by the rank of the gauge groups and the number of fundamentals on
each quiver node.

This paper is organized as follows: In section 2, we review the Hitchin system for superconformal case
and the regular singular solution is discussed;
In section 3, the irregular solutions to Hithcin's equation are introduced; In section 4,
the singularity types needed to describe $A$ type $SU(2)$ quiver gauge theory are written down; In section 5, we
generalize to $SU(N)$ case. Finally, we give the
conclusion and discuss some future directions. In appendix I, a detailed calculation about the
local dimension of Hitchin's moduli space around an irregular singularity and several simple
regular singularities is given and we show that it matches the dimension needed for the gauge theory.

\section{Review of Hitchin's Equation and $N=2$ Superconformal Field Theories}
A large class of four dimensional $N=2$ superconformal gauge theories can be engineered as the
six dimensional $(0,2)$ $A_{N-1}$ SCFT compactified on a Riemann surface with or without marked points.
In the case of marked points, defects are turned on on those marked points.
Hitchin's equation is argued to be the BPS equation for the compactification. This may be
seen by further compactifying four dimensional theory on a two dimensional torus, the final theory is
a two dimensional sigma model with certain target space. To identify the target space,
the compactification can be done in another order:  six dimensional theory is first compactified
on a torus and then on a Riemann surface with or without marked points. In the first step
of compactification, we get a four dimensional $N=4$ $SU(N)$ theory; In the second step of
compactification, we need to twist the theory to preserve some unbroken supersymmetry on
curved space time. There are three different kinds of twists for $N=4$ theory, it turns
out what is relevant for us is the so called Geometric Langlands (GL) twist \cite{GL}, and after this twist,
the BPS equation on Riemann surface is the Hitchin's equation and
 surface operators is inserted on marked points \cite{witten2, gukov}. The moduli space of solutions to Hitchin's equation is the target space for the two dimensional sigma model. Comparing with two types of compactification, we may conclude
that the BPS equation governing the six dimensional theory on Riemann surface is the Hitchin's equation, and
the description of the defects of the six dimensional theory is the same as the description of
surface operator in four dimensional gauge theory. This can also be seen by noting that in twisting the six dimensional theory, we get a one form gauge field and a one form scalar field on Riemann surface,
the same set of fields appear in GL twist of $N=4$ theory on a curved Riemann surface.

Hitchin's equation is studied extensively in \cite{GL}. Hitchin's equation with regular singularity on
Riemann surface $\Sigma$ is also studied by \cite{witten2, gukov} (called tamed ramification in Geometric Langlands Program). We review regular singularities in this section and leave the irregular singularity to next section.

Let's pick $SU(N)$ gauge group and write $g$ for its lie algebra, $t$ the lie algebra of the maximal torus $T$.
Hitchin's equation is
\begin{eqnarray}
F-\phi\wedge\phi=0\nonumber\\
D\phi=D*\phi=0,
\end{eqnarray}
where $F$ is the curvature of the connection $A_\mu$ of a vector bundle defined on Riemann surface $\Sigma$,
$\phi$ is the one form called Higgs field. The local behavior of conformal invariant solution to Hitchin's equation with regular singularity is (we consider one singularity here, multiple singularities can be studied similarly)
\begin{eqnarray}
A=\alpha d\theta +...\nonumber\\
\phi=\beta {dr\over r}-\gamma d\theta+...
\end{eqnarray}
where $\alpha, \beta, \gamma \in t$ (more precisely, $\alpha$ takes value in maximal torus $T$) and
less singular terms are not written explicitly; $z=re^{i\theta}$ is the local holomorphic coordinate.
The  moduli space of Hitchin's equation with the above behavior around the singularity is denoted as $M_H(\Sigma, \alpha, \beta,\gamma)$. $M_H(\Sigma, \alpha, \beta,\gamma)$ is a hyperkahler manifold and has a family of complex structures parameterized by $CP^1$.
These complex structures generically depend on the complex structure of the Riemann surface with some exception
which will be important to us later. In one distinguished complex structure I,
each point on moduli space represents a Higgs bundle on Riemann surface;
the complex structure modulus is $\beta+i\gamma$, and the kahler modulus is $\alpha$.
In the same complex structure, there is Hitchin's fibration which can be identified as
the Seiberg-Witten fibration. In the study of Seiberg-Witten curve of four dimensional theory,
only complex structure of the moduli space matters. since the residue of the Higgs field is $\sigma={1\over 2}(\beta+i\gamma)$ which determines the complex structure of the moduli space, we will focus
on the coefficient of Higgs field.

Assume $\alpha$ is regular with a Levi subgroup $L$, then the residue of Higgs field is taking value
in the lie algebra of the parabolic group determined by $\alpha$ modulo an element of the lie algebra $\textbf{n}$
of the unipotent radical of the parabolic group.
The massless theory corresponds to $\beta, \gamma \rightarrow 0$, in this
case, the above solution is not trivial, less singular terms can be added, and one can
show that the Higgs field has a simple pole with residue in $\textbf{n}$. Interestingly, the nilpotent
element of $sl_N$ is classified by Yang tableaux with total boxes N, so we can label the singularity
by the Yang tableaux, this is in agreement with the
result by studying Seiberg-Witten curve of $N=2$ SCFT \cite{Gaiotto1}. One can also show that Hitchin's fibration
is the same as the Seiberg-Witten fibration \cite{dan}. The massive theory is derived
by deforming to nonzero $\beta, \gamma$ regular with $L$, the form of $\beta+i\gamma$ is also determined by the same Yang tableaux.

The local moduli space of Hitchin's equation is described by the adjoint orbit ${\cal O}_i$ of the complex
lie algebra $sl(N,c)$ (the relation of the adjoint orbit
to moduli space of Nahm's equation can be found in \cite{kro1,kro2}) . The nilpotent orbit is used to describe the massless theory while the semi-simple
orbit is used to describe the mass-deformed theory. The nilpotent orbit is classified by the Young tableaux
$[n_1,n_2,...n_r]$ (for an introduction to nilpotent orbit, see \cite{nil}),
and the mass-deformed theory can be also read from  Young tableaux: there are a
total of $n_1$ mass parameters and the degeneracy of each mass parameter is equal to the number of
boxes on each column . There are only $n_1-1$ independent mass parameters because of the traceless condition.The dimension of the local moduli space is equal to the dimension of the
orbit ${\cal O}_i$:
\begin{equation}
N^2-\sum r_i^2, \label{local}
\end{equation}
where $r_i$ is the height of $i$th column of the Young tableaux. The Hitchin's moduli
space on the sphere can be modeled as the quotient
\begin{equation}
{{\cal O}_1\times{\cal O}_2...\times{\cal O}_m / G},
\end{equation}
where $G$ is the complex gauge group, and the total dimension is the sum of the local dimension minus
the dimension of the gauge group. The minimal nilpotent orbit has partition $[2,1,...1]$ and the mass-deformed theory has only one mass parameter, we call this kind of singularity as the simple regular singularity, the local
dimension of moduli space is $2N-2$ using (\ref{local}). The maximal nilpotent orbit
has partition $[N]$, and its dimension is $N^2-N$, we call it full regular singularity.

A large class of four dimensional $N=2$ superconformal field theories can be constructed by putting
together different punctures on Riemann surface, most of them do not have the conventional lagrangian description.
It is interesting to compare the $UV$ parameters of the gauge theory with known lagrangian description and
the parameters needed to define Hitchin's moduli space. Let's consider $A$ type
quiver gauge theory with gauge group $\sum_{i=1}^n SU(k_i)$ with $k_1<k_2<...<k_r=..=k_s>k_{s+1}>...>k_n$
with $k_r=...=k_s=N$. The
matter contents are the bi-fundamental hypermultiplet between the adjacent gauge groups and
the fundamentals $d_\alpha$ on each node to make gauge theory conformal, indeed
\begin{equation}
d_\alpha=2k_\alpha-k_{\alpha-1}-k_{\alpha+1}.
\end{equation}

Hitchin's system involves $n+1$ simple regular singularities and two
generic regular singularities on Riemann sphere. The two generic regular singularities are used to describe two quiver
tails, we study left quiver tail $SU(k_1)-SU(k_2)-...-SU(k_r)$ and right quiver can be treated similarly. It is
associated with a Young tableaux with partition $[n_1,...n_r]$, where $n_\alpha=k_\alpha-k_{\alpha-1}$,
there are $n_1-1$ mass parameters from this singularity. We also have $\sum_{\alpha=1}^r d_{\alpha}=k_1=n_1$.

Let's compare the UV parameters of the gauge theory with the parameters for the Hitchin system.
The UV parameters of the gauge theory are the dimensionless gauge couplings, mass parameters
for the bi-fundamental and fundamental hypermultilplets. The parameters for the Hitchin system
are the complex structure of the Riemann surface and the local parameters around the regular
singularities.   There are $n$ dimensionless gauge coupling constants,
and these are represented by the complex structure of punctured Riemann sphere, since there are $n+3$ punctures
on the sphere and the Riemann surface has $n$ complex structure moduli. There are $n-1$ mass parameters for bi-fundamental hypermultiplets and they are encoded in the parameters of $n-1$ regular simple singularities; For
the left quiver tail, there are a total of $n_1$ fundamental fields and $n_1$ mass parameters,
these are described  by a generic singularity and a regular simple singularity: the generic regular singularity
has $n_1-1$ mass parameters and the simple regular singularity has one. The same analysis
applies to the right quiver, so all the parameters are nicely encoded in the Hitchin's system. For
the IR behavior, the Seiberg-Witten fibration is identified with Hitchin's fibration, one can check
that the dimension of the base of Hitchin's fibration is the same as the dimension of
Coulomb branch of gauge theory.

Finally, we want to stress that the UV parameters enter into the Hitchin's system in different
ways. The Hitchin's moduli space is a hyperkahler space which is described by fixing the coefficient
of the simple pole, these coefficients describes the complex structure and kahler structure of
Hitchin's moduli space, and these parameters are identified with the mass parameters of the gauge theory, so
the hyperkahler structure depends on the mass parameters. On the other hand, the UV gauge coupling constants are identified with
the complex structure of the Riemann surface, but some of the complex structures of the Hitchin's
moduli space does not depend on the complex structure of Riemann surface, so the hyperkahler structure
is independent of the gauge coupling constant. This has important effect when we consider general
$N=2$ gauge theories in the following sections.

As we discussed in the introduction, it seems that the Riemann surface encodes all the information
about the gauge theory; On the other hand, two dimensional conformal field theory is
naturally defined on the Riemann surface with punctures. AGT \cite{Gaiotto2} found the
surprising relation between the Nekrasov partition function and conformal blocks
of Liouville theory; The relation
is extended to asymptotical free cases and $SU(N)$ conformal theory \cite{Gaiotto3, toda}. These relations
 have a lot of extensions and went through a lot of checks
\cite{ext1,ext2,ext3,ext4,ext5,ext6, che1,che2,che3,che4,che5,che6,che7,che8,che9,che10,che11,che12,che13,che14,che15,che16,che17,che18,che19}. One can also
compare the expectation values of  Wilson-t'hooft loops and surface operators
with correlation function of 2d CFT \cite{loop1,loop2,loop3,loop4,loop5,loop6,loop7}. AGT relation can be understood from matrix model \cite{mat1} and there are also a lot of developments along this line \cite{mat2,
mat3,mat4,mat5,mat6,mat7}. AGT is explained by
using $\Omega$ deformation and branes \cite{nek5} and M theory is also
useful in understanding the AGT relation \cite{M1, M2}. See also the development in understanding gauge theory side
\cite{SO,web, N1, dan2, yuji2, stony1, stony2}. Supergravity dual of the generalized quiver gauge theories
is found in \cite{malda}.

\section{Irregular Solutions to Hitchin's Equation}
Hitchin's equation is
\begin{eqnarray}
F-\phi\wedge\phi=0\nonumber\\
D\phi=D*\phi=0.
\end{eqnarray}
We want to find local solutions to Hitchin's equation and use it
to describe $N=2$ asymptotical free theory. In last section, it is shown
that regular singular solution is used to describe SCFT.
In the non-conformal case, the new feature is that
dimensional dynamical generated scale $\Lambda$ is included in UV parameters.
This UV parameter should enter into the description of the
Hitchin's system. In the conformal case, all the gauge coupling
constants are dimensionless, they are identified with the
complex structure moduli of the Riemann surface. They are not entering
into the description of the hyperkahler structure of the moduli
space of Hitchin's equation. The mass parameters do enter into
the description of the hyperkahler structure, they are the
parameters of the coefficient of regular pole, and the
parameters of the regular pole has topological meaning.
One of reason why the UV parameters have different description
is that Hitchin's moduli space is the Coulomb branch of the
three dimensional theory derived by compactifying four dimensional
theory on a circle, but the three dimensional gauge coupling
is not conformal any more and therefore the four dimensional
conformal gauge couplings do not have the topological meaning
on Hitchin's moduli space. We have to describe $\Lambda$ in a non-topological
way as we describe the dimensionless gauge coupling.
Since $\Lambda$ is dimensional, we can not describe
it as the dimensionless complex structure moduli. The dimensional
field in the Hitchin's system is the Higgs field, so $\Lambda$
should enter into the definition of the Higgs field.
The simple pole case is not enough, since the parameter has
the topological meaning. Then we conclude that
higher order singularity is needed to describe asymptotical free theories.
In fact, the parameters for higher order singularity
are shown to not have the topological meaning \cite{wild}.

Hitchin's equation does have solutions with irregular singularities.
It is the purpose of this paper to identify what kind of irregular singularities are
needed to describe four dimensional $N=2$ gauge theories. Consider an irregular singularity at the origin, Hitchin's equation is schematically $d\Phi+\Phi^2=0$ for $\Phi=(A,\phi)$, so for solutions
singular than ${1\over z}$, they are not compatible unless the
solution is abelian, namely, they are taking values in a Cartan
subalgebra. See the detailed explanation in \cite{wild}, we give a short
review in the below.

Introduce local coordinate $z=re^{i\theta}$, and let $t$ denote the lie algebra
of a maximal torus $T$ of the compact lie group G (we take G as $SU(N)$ in this paper) and $t_C$
its complexification. We pick $\alpha\in t$ and $u_1,...u_n \in t_C$,
and consider the solution
\begin{eqnarray}
A&=&\alpha d\theta +....  \nonumber\\
\phi&=&{dz\over 2}({u_n\over z^n}+{u_{n-1}\over z^{n-1}}+...+{u_1\over z})
+{d\bar{z}\over 2}({\bar{u}_n\over \bar{z}_n}+{\bar{u}_{n-1}\over \bar{z}^{n-1}}
+...+{\bar{u}_1\over \bar{z}}+...).
\end{eqnarray}
Let's first assume that $u_n$ is regular and semi-simple, and $u_1=\beta+i\gamma$; When $n=1$,
this solution is reduced to the simple pole case.

Let's denote the moduli space of the solution as $M_H$. The moduli space
has the hyperkahler structure and have three distinguished complex structures.
In complex structure I,  each pair of
solutions $(\phi, A)$ represents a Higgs bundle. The holomorphic structure of the bundle $E$ is
defined by using the $(0,1)$ part of the gauge field $A$. The $(1,0)$
part $\Phi$ of $\phi$ is a holomorphic section of $ad(E)\bigotimes K_C$.
Explicitly, do a complex conjugation using $r^{i\alpha}$, the operator
$\bar{\partial}_A=d\bar{z}(\partial_{\bar{z}}+A_{\bar{z}})$ reduces to the
standard one $d\bar{z}\partial_{\bar{z}}$. With this trivialization, The holomorphic part of Higgs field is
\begin{equation}
\Phi={dz\over 2}({u_n\over z^n}+{u_{n-1}\over z^{n-1}}+...+{u_1\over z}).
\end{equation}
There is also a Hitchin's fibration and the spectral curve
\begin{equation}
\det(x-\Phi(z))=0.
\end{equation}
This Hitchin's fibration is identified with Seiberg-Witten fibration and the spectral
curve is the Seiberg-Witten curve.

In another complex structure J, We study the $G_c$ valued complex connection
${\cal A}=A+i\phi$, which is flat by using  Hitchin's equation.
The connection ${\cal A}$ can be put in the form
\begin{equation}
{\cal A}_z=({u_n\over z^n}+{u_{n-1}\over z^{n-1}}+...+{u_2\over z^2})-i{\alpha-i\gamma\over z}
\end{equation}
The connection can be put in a standard form:
\begin{equation}
{\cal A}_z={T_n\over z^n}+{T_{n-1}\over z^{n-1}}+....+{T_1\over z}.
\end{equation}

For such irregular connection, the monodromy is not just determined by $T_1$, the stokes matrix
is needed to describe the so-called generalized monodromy. The dimension of the local moduli space
is
\begin{equation}
dim(M_H)=(n)(dim(G_c)-r),
\end{equation}
here $r$ is the rank of the gauge group and $T_k$ are chosen in a regular semi-simple orbit,
this dimension can be derived by counting the parameters for the generalized monodromy.
In defining the moduli space, the matrices $T_n,...T_1$ are fixed,
so when  Hitchin's system is used to describe four dimensional gauge theory, these parameters
are interpreted as the parameters like the masses, dynamical generated scale.
Hitchin's moduli space is identified with the Coulomb branch moduli space.
We should emphasize that the moduli space does not depend on $T_n...T_2$ \cite{wild} in complex structure J, so we
should identify dynamical scale with the parameters in $T_n...T_2$ as we argued at the beginning of
this section.

What happens when the leading order coefficient is not regular-semisimple? If $u_n$ is
semi-simple but not regular, the analysis is essentially the same as described in section 6 of \cite{wild}.
When $u_n$ is nilpotent, the solution can be reduced to the solution with leading order coefficient semi-simple. This is in contrast to the simple pole solution, in that case, when the residue is nilpotent, we have new solutions
to the Hitchin equation.

The solution with nilpotent leading order coefficient plays an essential role in describing $N=2$
gauge theories. Let's consider the case ${\cal A}_z=T_n/z^n+...,$ with $T_n$ nilpotent. We
take $G_c=SL(2,C)$ for an example. ${\cal A}_z$ can be written as
\begin{equation}
{\cal A}_z=\left(\begin{array}{cc}
a&~z^{-n}b\\
c&~-a\end{array}\right),
\end{equation}
where $ a,c $ have poles at most of order $n-1$ at $z=0$  and $b$ is regular.
Now by a gauge transformation $g=\left(\begin{array}{cc}1&~0\\f(z)&~1\end{array}\right)$,
we can set $a=0$ by choosing appropriate $f(z)$, the connection becomes
\begin{equation}
{\cal A}_z=\left(\begin{array}{cc}
0&z^{-n}b\\
z^{-k}\tilde{c}&0 \end{array}\right),
\end{equation}
where $\tilde{c}$ is regular and $k<n$. If $n-k>2$, we can make a further gauge transformation with
$g=\left(\begin{array}{cc}z^{1\over2}&~0\\0&~z^{-1\over 2}\end{array}\right)$ and follow a similar
gauge transformation to take the connection back to off-diagonal form to reduce $n$ and $n-k$.
The only new possibility is then $n=k$ or $n=k+1$, if $n=k$, we are back to the case with $T_n$
regular semi-simple. If $n=k+1$, we take a double cover of a neighborhood around the singular
point. We introduce a new coordinate $z=t^2$, then the solution is reduced to previous situation with
a gauge transformation $g=\left(\begin{array}{cc}t^{1\over2}&~0\\0&~t^{-1\over 2}\end{array}\right)$.
Write ${\cal A}={\cal A}_t dt$, and then ${\cal A}_z={\cal A}_t/2t$. The connection has the form
\begin{equation}
A_t=\left(\begin{array}{cc} 0&t^{-2n}b\\
t^{-2n}c&0\end{array}\right).
\end{equation}
$A_t$ is even under $t\rightarrow -t$, so $A_tdt$ is odd under $t\rightarrow -t$. So the
singular solution becomes
\begin{eqnarray}
A=0 \nonumber\\
\phi={dt\over 2}({v_{n-1}\over t^{2(n-1)}}+{v_{n-2}\over t^{2(n-2)}}+...+{v_1\over t^2})+c.c.
\end{eqnarray}
Now the leading order coefficient is regular semi-simple. It is useful to transform the solution to the original
coordinate $z$:
\begin{eqnarray}
A=0, \nonumber \\
\phi={dz\over 4}({v_{n-1}\over z^{n-1/2}}+{v_{n-2}\over z^{n-3/2}}+....+{v_1\over z^{3\over 2}})+c.c \label{nil}
\end{eqnarray}
To make this solution well defined, we need to make a gauge transformation $M$ in crossing the cut
on the $z$ plane
\begin{equation}
M=\left(\begin{array}{cc} 0&1\\
1&0\end{array}\right).
\end{equation}
We can also add the regular terms, the regular terms are of the form $v_kz^{k-1/2}, k\geq1$
to make the solution well defined, the regular singular term is missing here so this singularity
does not encode any mass parameter.

For instance, if $n=2$, the holomorphic part of the Higgs field (we will call the holomorphic part of the
Higgs field as Higgs field in later parts of this paper)is $\Phi_z={v_1 \over z^{3/2}}+{C\over z^{1/2}}+...$,
The spectral curve is $x^2=\phi_2(z)$, where $\phi_2(z)=Tr(\Phi_z)^2$ is the degree two differential on
the Riemann surface. The quadratic differential has the form
\begin{equation}
\phi_2(z)=Tr (\Phi_z)^2={q^2\over z^3} +{U\over z^2}+{M\over z}+...
\end{equation}
where $v_1=diag(q,-q)$ and $C=diag(a,-a)$. The parameter $U$ depends on $a$. This parameter is
identified with the coulomb branch of the gauge theory, since $\phi_2(z)$ is a degree 2
meromorphic connection, according to Riemann-Roch theorem, this pole contribute two to the Coulomb branch, and
so it contributes four to the Hitchin's moduli space. The above method shows how to calculate
the local dimension of the moduli space in Hitchin's equation. We expand the spectral curve around
the singularity, and find the maximal pole of degree $i$ differential which depends on regular
term, and then sum up the contributions from degree $2$ differential to degree $N$ differential
for $SU(N)$ case, this gives the local dimension of the base, the total dimension of local
Hitchin's moduli space is twice of the base.

The form of the gauge field and Higgs field can be derived in another way. Let's take $n=2$
for an example. Suppose the Higgs field takes the following form
\begin{equation}
\Phi(z)={1\over z^2}\left(\begin{array}{cc} 0&1\\
0&0\end{array}\right)dz+{1\over z}\left(\begin{array}{cc} a&b\\
-c&d\end{array}\right)dz+.....
\end{equation}
One can calculate the eigenvalues of $\Phi$ which are ${1\over z^{3/2}}(\Lambda, -\Lambda)$, and
the ${1\over z}$ term is missing because the leading order term is multivalued while ${1\over z}$
term is single valued, so we do not have any mass deformation
for this type of singularity, this recovers (\ref{nil}). The local dimension of the moduli
space can be derived by noting that the leading order coefficient belongs to nilpotent orbit with dimension 2
and the regular singularity coefficient is in a semi-simple orbit also with dimension 2,
so the local dimension is $2+2=4$.

More generally, the connection ${\cal A}_z$ is an $N\times N$ matrix-valued function with a possible pole
at $z=0$. It has N possibly multiple eigenvalues $\lambda_i$. The eigenvalues behave for small z as $z\sim z^{-r_i}$,
with rational number $r_i$. Tame ramification is the case that all $r_i$ are equal to or less than 1. We call
completely wild ramification if $r_i>1$ for all i. The general case is a mixture of these two possibilities.
Following $SU(2)$ case, Let's consider second order irregular singularity with leading order coefficient
nilpotent
\begin{equation}
\Phi(z)={A_1\over z^2}dz+{A_0\over z}dz+....,
\end{equation}
where $A_1$ is the matrix in the nilpotent orbit ${\cal O}_1$ labeled by Young tableaux with partition $[2,1,1..,1]$,
we take $A_1$ as the matrix with standard Jordan form; $A_0$ is in a regular semi-simple orbit ${\cal O}_0$ (the eigenvalues of $A_0$ are all distinct). One can
calculate the eigenvalue of the Higgs field, it has the form
\begin{equation}
\Phi={1 \over z^{1+{1/N}}} diag(1,\omega,\omega^2...\omega^{N-1})dz,
\end{equation}
where $\omega^N=1$; Similar as $SU(2)$ case, $1\over z$ term is missing. The local dimension of the Hitchin's moduli space is equal to the sum of the dimension of
the orbit ${\cal O}_1$ and ${\cal O}_0$,
\begin{equation}
d=2N-2+N^2-N=N^2+N-2, \label{d}
\end{equation}
Notice that this equals to the contribution of a simple regular singularity and a full regular singularity
with partition $[n]$. This irregular solution is useful to us since there is only one parameter in the
irregular part and this can be identified with the dynamical scale and there is no mass deformation, so this
irregular singularity is useful for the pure $N=2$ super Yang-Mills theory, this will be confirmed in
later sections. This solution may be seen as the minimal irregular singularity.

We also want to have irregular solutions allowing some mass deformation. It turns out that
the Higgs field with following eigenvalues is useful (after diagonalization, here we use $n$ instead of $N$ to denote the
rank of gauge group)
\begin{equation}
\Phi={1\over z^{1+{1\over n-k}}}diag(0,...0,\Lambda, \Lambda \omega,...\Lambda\omega^{n-k-1})dz+{1\over z}diag(m_1,m_2,..m_k,m_{k+1}, m_{k+1}...m_{k+1})dz+..., \label{higgs}
\end{equation}
where $\omega=e^{-2\pi i\over n-k}$ and the sum of mass vanishes so we have $k$ independent
mass parameters.

This form needs a small change for $k=n-1$, in this case, the Higgs field has the form
\begin{equation}
\Phi={1\over z^{2}}diag(\Lambda,\Lambda,..,\Lambda, -(n-1)\Lambda)dz+{1\over z}diag(m_1,m_2,..m_n)dz+..., \label{higgs1}
\end{equation}
$\sum_{i=1}^{n}m_i=0$. A special case is when $n=2$, the leading order
coefficient is regular semisimple.

The solution (\ref{higgs}) is well defined only when
 a gauge transformation is made on crossing the cut on z plane:
\begin{equation}
M=\left(\begin{array}{ccccc}
0&.&.&.&0\\
.&0&.&.&0\\
.&.&.&.&0\\
.&.&.&0&0\\
0&.&.&.&v_{n-k} \end{array}\right),
\end{equation}
where $v_{n-k}$ is the $(n-k)\times (n-k)$ matrix
\begin{equation}
v_{n-k}=\left(\begin{array}{cccccc}
0&1&0&...&0&0\\
0&0&1&...&0&0\\
...&...&...&...&...&...\\
0&0&0&...&0&1\\
1&0&0&...&0&0 \end{array}\right). \label{mat}
\end{equation}
Let's first consider the simple case $k=0$, in which the simple pole term is forbidden. We can also
add the regular terms, however, the regular term must take a form so that the Higgs field is well defined
in crossing the cut in the z plane. The Higgs field takes the following form
\begin{equation}
\Phi={\Lambda\over z^{1+1/n}}diag(1,\omega,...\omega^{n-1})+\sum_{d=1}^{n-1}{a_d\over z^{1-d/n}}diag(1,\omega^{-d}..\omega^{-dj},...)+...
\end{equation}
One can check that the Higgs field is well defined using the gauge transformation (\ref{mat}) in
crossing the cut.
Since $(1,\omega,...\omega^{n-1})$ are the roots of the equation $x^n-1=0$, the equation factorizes as
$x^n-1=\sum_{i=0}^{n-1}(x-\omega^i)$, expanding the last equation, we have the relation $\sum_iw^i=0$,
$\sum_{i\neq j}\omega^i\omega^j=0$, and $\sum_{i\neq j\neq k} \omega^i\omega^j\omega^k=0$, etc, the only
nonvanishing combination is $\prod w^0w...w^{n-1}=(-1)^{n-1}$. We also have the relation $\sum_{j=0}^{n-1}w^{-dj}=0$
for any $d$.

Calculating the determinant
\begin{equation}
det(x-\phi(z))=x^n-\sum_{i=2}^n\phi_i(z)x^{n-i}.
\end{equation}
The leading singular behavior of the coefficient $\phi_i(z)$ can be derived by
expanding the determinant. For $\phi_2$, one may wonder it
has the fractional power, but this is not the case, since the coefficient of ${1\over z^{2+{2\over k}}}$ is $\sum_{i\neq j} w^iw^j=0$. Let's calculate the leading order term depending on the regular coefficient, it has the form
\begin{equation}
\phi_2(z)={1\over 2}\sum_{i\neq j}\omega^i\omega^{-dj}{1\over z^{2+{(1-d)/n}}}=\sum_i\omega^i\omega^{-di}{1\over z^{2+{(1-d)/n}}}
=\sum_i\omega^{(1-d)i}{1\over z^{2+{(1-d)/n}}}.
\end{equation}
This term is nonzero only in the case $d=1$. So $\phi_2(z)={c\over z^2}+..$.

The calculation can be
extended to the other coefficient $\phi_i$, there is no term which only depends on the singular term of
the Higgs field except for $i=n$, the leading order terms depending on the regular terms are
\begin{equation}
\phi_i(z)=C\sum_{n_1\neq n_2..n_{i}}\omega^{n_1}...\omega^{n_{i-1}}\omega^{-dn_i}{1\over z^{i+(i-d)/n}} =C\sum_j\omega^{(i-d)j}z^{i+(i-d)/n} \label{pure}.
\end{equation}
We select $n-1$ terms from the irregular term and one regular term, the only vanishing term is when $d=i$,
so $\phi_i=C{1\over z^i}+...$.

For the coefficient $\phi_n(z)$, there is a term depending on $\Lambda$, it has
the form
\begin{equation}
\phi_n(z)={\Lambda^n\over z^{n+1}}+C{1\over z^n}.
\end{equation}
The contribution of this singularity to the Coulomb branch is
\begin{equation}
2+...n={n^2+n-2\over 2}.
\end{equation}
This is the same as we calculate by counting the dimension of the adjoint orbit for
the coefficient on the singular part (\ref{d}).

Let's define the local covering coordinate $z=t^n$, the Higgs field has the form
\begin{equation}
\phi(t)={1\over t^2}diag(1,\omega,...\omega^{n-1})dt+....
\end{equation}
To make this field well defined, we can not turn on the regular singular term. Notice
that this has the familiar form with leading order coefficient semi-simple.

One can similarly study the spectral curve of the Higgs field (\ref{higgs},\ref{higgs1}), the term depending on the
regular term is $\phi_i(z)={C\over z^i}$, so the contribution to the coulomb
branch of this singularity is also ${n^2+n-2\over 2}$.
The difference here is that we also have the
terms with coefficient depending on the singular terms, i.e. mass parameters and dynamical scale.

To summarize, we have studied several types of irregular singularities: For the first one, the leading order
coefficient is regular semi-simple and this kind of singularity is studied extensively; if the leading order coefficient is semisimple but not regular, it can be treated similarly with some complication; if the leading order singularity is nilpotent, the solution can be transformed to a form as the first two by going to local
covering space.

\section{SU(2) Theory}
In this section, we will identify the corresponding Hitchin system for $N=2$ $SU(2)$ gauge theory. One important
clue for the Hitchin system is that its total complex dimension must be 2, so the base of the
Hitchin fibration is 1 and can be matched with the dimension of the Coulomb branch of $N=2$ SU(2) theory.
The corresponding Seiberg-Witten curves are calculated and they have the same form as derived in \cite{moore}.

Let's first recall the six dimensional construction of $SU(2)$ theory with four fundamentals: it is
represented by a $SU(2)$ Hitchin system defined on a Riemann sphere with four punctures, at each puncture
the Higgs field has the simple pole with the residue $diag(m,-m)$. It is useful to think this in the brane
picture \cite{witten3}, one $D4-NS5$ brane configuration is depicted in Figure 1a). This is a type IIA
construction. The NS5 branes
which extend in the direction $x^0, x^1,x^2,x^3,x^4,x^5$, are
sitting at $x^7,x^8,x^9=0$ and at the arbitrary value of $x^6$. The
$x^6$ position is only well defined classically. The D4 branes are
stretched between the fivebranes and their world volume is in $x^0,x^1,x^2,x^3$ direction;
These D4 branes have finite length in $x^6$ direction. We also have $D6$ branes which
extend in the direction $x^0,x^1,x^2,x^3,x^7,x^8,x^9$.

We can think that two punctures describe the behavior of two semi-infinite branes on the left, and two others used to describe semi-infinite branes on the right. The gauge coupling constant is identified with the complex structure moduli of the punctured sphere.
\begin{figure}
\begin{center}
\includegraphics[width=4in,]
{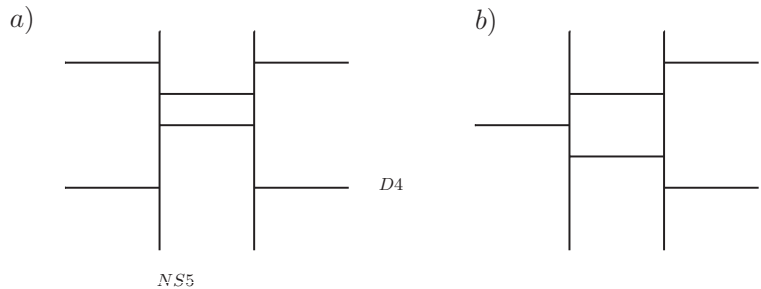}
\end{center}
\caption{a) Brane configuration of $SU(2)$ theory with four fundamentals; b)Brane configuration of $SU(2)$
theory with three fundamentals. }
\end{figure}

Next, let's consider the $SU(2)$ theory with $N_f=3$, the brane configuration is depicted in figure 1b).
The maximal number of punctures on the Riemann sphere are three since we do not have a dimensionless
gauge coupling, and there is at least one irregular singularity as we argued in last section about the
asymptotical free theory. Since the
right hand side of brane configuration is the same as $N_f=4$ theory, we expect that there are also
two simple punctures on the Riemann sphere; On the left hand side, there are only one semi-infinite D4 brane,
so one more irregular singularity is needed. Another important clue is that the coulomb branch is one dimensional, so the dimension of the Hitchin's space is $d=2$, the dimension of Hitchin's moduli space on a sphere with two simple punctures and one irregular puncture is
\begin{equation}
d=d_1+(2N-2)+(2N-2)-2(N^2-1)=2,
\end{equation}
where $N=2$ and $d_1$ is the local dimension of the irregular singularity. We have $d_1=4$, therefore
the local dimension of the irregular singularity is $4$. Let's look at
the irregular singularity (\ref{higgs},\ref{higgs1}) we discussed in last section,
they contribute to the coulomb branch dimension $d={{N^2+N-2}\over2}=2$, we have two choices $k=0$, or
$k=1$. There is  two parameters for $k=2$ and one for $k=1$. To see which one is
the correct one for $N_f=3$ theory, it is useful to check the parameters of the
gauge theory, there are three mass parameters for the three fundamentals and one dynamical generated
scale $\Lambda$, and two mass parameters are associated with two simple punctures and
the remaining parameters are associated with the irregular singularity, so we must select $k=1$,
and the Higgs field has the form around this singularity
\begin{equation}
\Phi={1\over z^2}\left(\begin{array}{cc}
\Lambda&0\\
0&-\Lambda \end{array}\right)dz+{1\over z}\left(\begin{array}{cc}
m_3&0\\
0&-m_3 \end{array}\right)dz+..., \label{simple1}
\end{equation}
and the Higgs field has the form at the other two simple punctures
\begin{equation}
\Phi={1\over z}\left(\begin{array}{cc}
m_i&0\\
0&-m_i \end{array}\right)dz+..., \label{simple}
\end{equation}
with $i=1,2$.

The Seiberg-Witten curve associated with this Hitchin system is
\begin{eqnarray}
det(x-\Phi(z))=0 \nonumber\\
x^2={m_1^2\over z^2}+{m_2^2\over (z-1)^2}+{U\over z(z-1)}+{2m_3\Lambda\over z}+\Lambda^2
\end{eqnarray}
We have used the conformal symmetry to put the simple punctures at $z=0, 1$ and the
irregular puncture at $z=\infty$; $U$ is the Coulomb branch parameter.

Next, let's consider $N_f=2$ case, there are two different brane configurations, which
is depicted in Figure 2.
\begin{figure}
\begin{center}
\includegraphics[width=4in,]
{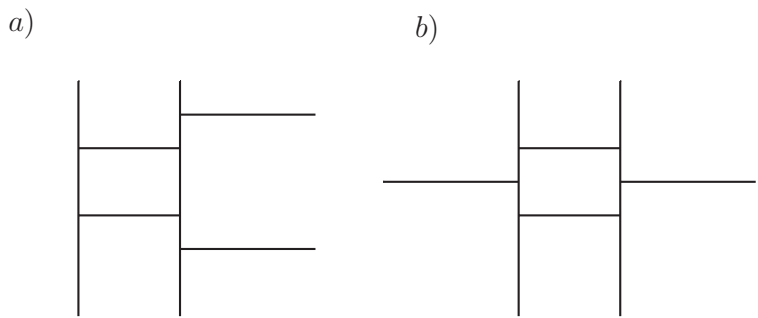}
\end{center}
\caption{a) Brane configuration of $SU(2)$ theory with two fundamentals; b) Another Brane configuration of $SU(2)$
theory with two fundamentals. }
\end{figure}

For the brane configuration in Figure 2a), two simple punctures are needed to describe two fundamentals on
the right-hand side.  There is no fundamental on the left hand side, so we need to
use the Higgs field ($\ref{higgs1}$) with $k=0$. We should point out that the number $k$ in our solution (\ref{higgs},\ref{higgs1}) has been identified with the number of fundamentals on the left hand side of the brane configuration. When $k=2$, the irregular singularity
becomes two simple regular singularities.

The Higgs field at the irregular puncture for $N_f=2$ case is
\begin{equation}
\Phi(z)={1\over z^{3/2}}\left(\begin{array}{cc}
\Lambda &0\\
0&-\Lambda \end{array}\right)dz+.....\label{simple2}
\end{equation}
The Higgs field at the other two regular punctures have the same form as in (\ref{simple}). The spectral
curve has the form
\begin{eqnarray}
det(x-\Phi(z))=0 \nonumber\\
x^2={m_1^2\over z^2}+{m_2^2\over (z-1)^2}+{U\over z(z-1)}+{\Lambda^2\over z}
\end{eqnarray}
We have put two simple regular singularities at $z=0,1$, and the irregular singularity at $z=\infty$.

For the brane configuration in Figure 2b), We have one fundamental on the left hand side and
one fundamental on the right hand side, the brane configuration is symmetric on both sides. We
have two punctures and they must be irregular to account for the dimension.
As we argued earlier, the solution corresponds to $n=2, k=1$ in (\ref{simple1}), so the Higgs field takes the form
(\ref{simple1}), the parameter $\Lambda$ must be same for two punctures and the mass parameters
are different.

The spectral curve takes the form
\begin{eqnarray}
det(x-\Phi(z))=0 \nonumber\\
x^2={\lambda\over z^4}+{m_1\Lambda\over z^3}+{U\over z^2}+{\Lambda m_2\over z}+\Lambda^2
\end{eqnarray}
We put the puncture at $z=0,\infty$.

We next consider  $N_f=1$, from the brane configuration in Figure 3a), we can conclude that
we have two irregular punctures, one of them takes the form (\ref{simple2}); the
other has the form (\ref{simple1}).
The spectral curve is
\begin{eqnarray}
det(x-\Phi(z))=0 \nonumber\\
x^2={\lambda^2\over z^3}+{U\over z^2}+{\Lambda m\over z}+\Lambda^2
\end{eqnarray}
punctures are put at $z=0, \infty$.
\begin{figure}
\begin{center}
\includegraphics[width=4in,]
{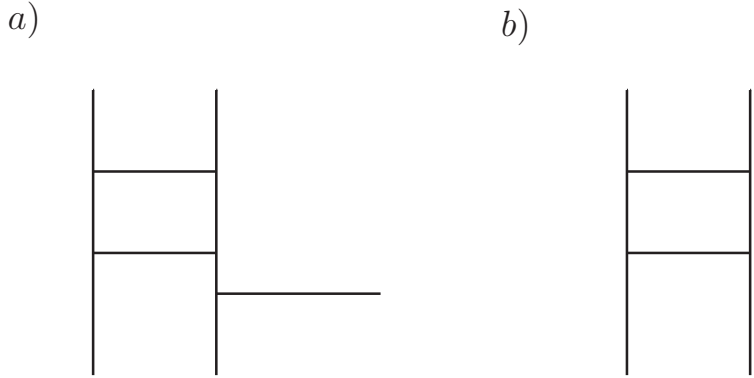}
\end{center}
\caption{a) Brane configuration of $SU(2)$ theory with one fundamentals; b) Another Brane configuration of pure $SU(2)$ theory  }
\end{figure}

We then consider the pure $N=2$ $SU(2)$ theory. The brane configuration is in Figure 3b). We have
two irregular singularities with the Higgs field taking the form (\ref{simple2}), the coefficient
for the leading order coefficient is the same for two punctures due to symmetry. The spectral curve
is
\begin{eqnarray}
det(x-\phi(z))=0 \nonumber\\
x^2={\Lambda^2\over z^3}+{U\over z^2}+{\Lambda^2\over z}
\end{eqnarray}
The Seiberg-Witten curve above is the same as derived in \cite{moore}.

The above analysis exhausted $N=2$ $SU(2)$ gauge theories with fundamental hypermultiplet.
 There are more choices for the Hitchin moduli space with total
dimension 2 though. For a singularity with regular semi-simple coefficient, the
local dimension is $2n$, where $n$ is the order of the singularity, this includes the
regular singularity. If the leading
coefficient is nilpotent, as we showed in last section, the dimension is also $2n$, if
the leading order singularity has the form $\Phi={v_n\over z^{n-1/2}}dz, n=2..$. We
have the following choices except those studied in this section:

i) One 3 order irregular singularity with regular semi-simple coefficient and one regular
singularity, we have two mass parameters associated with the residue of the regular singularity.
This might be related to  $A_2$ Argyres-Douglas fixed point \cite{ag,ag1}.

ii) One 3 order irregular singularity with the form (\ref{nil}) with $n=3$, and a simple singularity, we
suspect this is related to $A_1$ Argyres-Douglas fixe point

iii) One 4 order irregular singularity with regular semi-simple coefficient, we suspect this is also related
to $A_1$ Argyres-Douglas fixed point.

iv) One 4 order irregular singularity with  the form (\ref{nil}), we suspect this is associated
with the $A_0$ Argyres-Douglas superconformal fixed point.

There are some clues that the above conjecture might be true. Singular fibre is classified by Kodaira,
and Argyres-Douglas fixed point corresponds to singular fibre of type $A_2$, $A_1$ and $A_0$. According to
the result by Boalch \cite{boalch}, case i) can be associated with the affine dynkin diagram of $A_2$, case ii) is
associated with affine dynkin diagram of $A_1$ and case iv) is associated with dynkin diagram of $A_0$.
There is another hint about our conjecture: For $A_2$ Argyres-Douglas fixed point, there are two deformation
parameters, and we have two deformation parameters for the Hitchin system i), one from the order
3 singularity and the other from regular singularity; Case ii) and case iii) both have one mass
parameter and  match the deformation parameter for $A_1$ theory;
Case iv) does not have mass parameter which match the deformation parameter of
$A_0$ theory.

It is natural to generalize the above study to the linear quiver with only $SU(2)$ gauge groups.
For the superconformal case with $n$ $SU(2)$ gauge group, there are
a total of $n+3$ regular singular punctures on the sphere, with $n-1$ punctures
to account for the flavor symmetry of the bi-fundamental and 2 puncture for the two fundamentals on the
far left, and 2 punctures to account for the bi-fundamental on the far right. If the quiver is not conformal,
only the gauge groups at both ends are not conformal; we need to replace the two simple punctures
with the irregular puncture based on solution (\ref{higgs},\ref{higgs1}) with $n=2, k$, where $k$ is the
number of fundamentals on the end; if $k=2$, we still have two simple regular punctures. There are a total of $n+1$
punctures on the sphere if $k<2$ on both ends, and we have $n-2$ conformal gauge groups ; the
punctured sphere has a total of $n-2$ moduli and these moduli are identified with the UV gauge couplings of
the $(n-2)$ conformal gauge group.
\begin{figure}
\begin{center}
\includegraphics[width=4in,]
{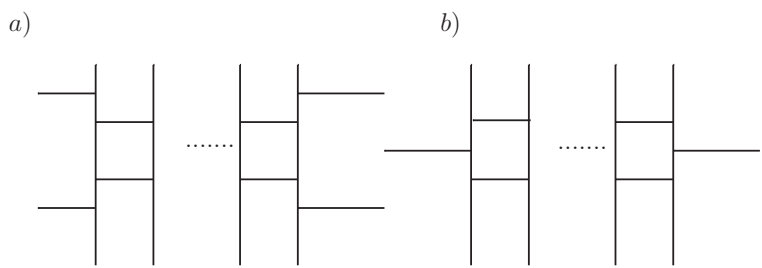}
\end{center}
\caption{a) Brane configuration of conformal $SU(2)$ quivers; b) A brane configuration with non-conformal gauge group   }
\end{figure}

In conclusion, to describe $SU(2)$ linear quiver, besides the regular singularity, we also need to turn on
two types of irregular singularities (\ref{simple},\ref{simple2}). One can similarly studied the different degeneration limits and we will get
generalized quiver as in \cite{Gaiotto1}.
For example, let's consider the quiver in Figure 5a). There are three $SU(2)$ gauge groups and only the middle
one is conformal. The weakly coupled limit of quiver 5a) is described by the degeneration limit of the
Riemann sphere with four punctures in Figure 5b): we have two simple punctures and two irregular punctures described
by boxes. Figure 5c) describes another degeneration limit of the same Riemann sphere, after the complete degeneration limit, we get a theory which is described by Hitchin's equation with two irregular singularities and one regular singularity, this is depicted in
Figure 5d). This theory has two dimension two operators in Coulomb branch and it is a linear quiver with two $SU(2)$ gauge group. The dual quiver is a generalized quiver.
\begin{figure}
\begin{center}
\includegraphics[width=4in,]
{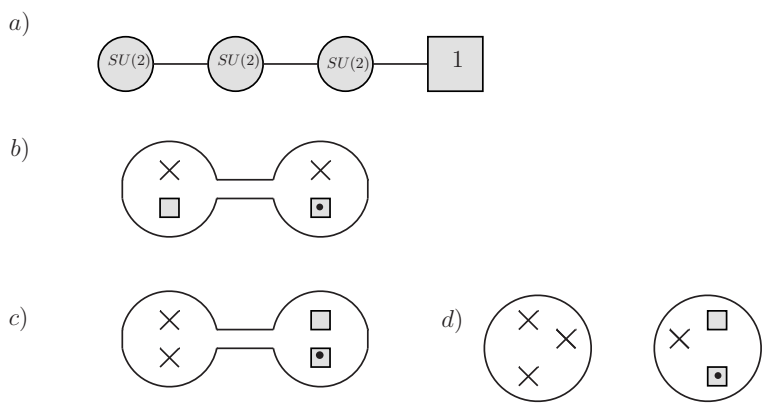}
\end{center}
\caption{a) A nonconformal $SU(2)$ quiver with three gauge groups;
 b) The degeneration limit of the Riemann sphere corresponding to quiver a), the regular singularity is
 represented by cross, and the irregular singularity is represented by box; c) Another
 degeneration limit of the same Riemann sphere; d) After complete degeneration, we
 get a new $SU(2)$ theories with two irregular singularity and a simple singularity.   }
\end{figure}

More generally, there are two kinds of singularities for $SU(2)$ Hitchin system:  order n
singularity with coefficient semi-simple or nilpotent. We can engineer gauge theories by putting these singularities
on Riemann surface. Fortunately, for the known gauge theory with conventional lagrangian
description, we need the simplest ones, namely, order two or order one singularity.

\section{SU(N) Theory}
Let's now generalize the analysis of $SU(2)$ theory to $SU(N)$ case. We first consider a single $SU(N)$ gauge group with $N_f$ fundamental hypermultiplets. The conformal case with $N_f=2N$ is reviewed in section two.
The theory is described by six dimensional $A_{N-1}$ theory compactified on a sphere with four punctures:
two simple regular punctures and two full regular punctures.

The brane configuration corresponds to put N fundamentals to the left and N fundamentals to the right as
in the Figure 1a). We can think that one simple puncture and one full puncture are needed to describe
$N$ fundamentals on each side. The contribution of these two punctures to Coulomb branch parameters
is $N-1+{1\over 2}(N^2-N)={N^2+N-2\over 2}$, where $N-1$ is the contribution of the simple regular puncture
and ${1\over 2}(N^2-N)$ is the contribution of the full regular puncture.

The brane construction is still very useful to consider asymptotical free theory; In analogy with
$SU(2)$ theories, let's first consider $2N>N_f\geq[{1\over 2} N]$, in this case
we can put N fundamentals on the right and $N_f-N$ fundamentals on the left, and  a simple and a full regular punctures are needed to describe the $N$ fundamentals on the right.
An irregular puncture is needed to describe the fundamentals on the left.
To get the correct number of coulomb branch moduli, the contribution of the irregular singularity to
Coulomb moduli space must be ${1\over 2}(N^2+N-2)$; We do have a class of irregular singularity with
this number in (\ref{higgs},\ref{higgs1}).  From the analysis of $SU(2)$ theory, we may want to select
the solution with $n=N, k=N_f-N$, an important check is that the flavor symmetry on the $N-N_f$ fundamentals
on the left hand side is $U(k)$. The regular singular part of the irregular singularity has
the partition $(k+1,1,1,...1)$, which do describe $U(N)$ flavor symmetry.

We now have the clue to describe $SU(N)$ theory with any number of fundamentals. We can decompose
$N_f=k_{-}+k_{+}$ and $k_{-}\leq k_{+}\leq N$, namely, we put $k_{-}$ semi-infinite D4 branes to the left
and $k_{+}$ semi-infinite D4 branes to the right. See the brane configuration in Figure 6. If $k_{+}<N$,  two irregular singularities are needed,
and the local solutions are of the form (\ref{higgs},\ref{higgs1}) with $n=N, k=k_{-}$ and $n=N, k=k_{+}$; if
$k_{+}=N$, we have two regular punctures and one irregular puncture with $n=N, k=k_{-}$. We also
need to set the coefficient $\Lambda$ at the irregular singularities equal.

The Seiberg-Witten curve is
derived from the spectral curve of the Hitchin system. Notice that, there are more than one description for the same
$SU(N)$ theory with $N_f$ fundamentals. The different Hitchin moduli spaces corresponding to different
decomposition of $N_f$ are isomorphic! \cite{boalch, moore}
\begin{figure}
\begin{center}
\includegraphics[width=4in,]
{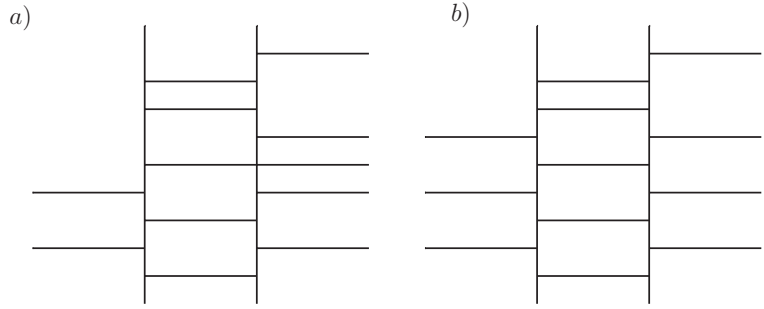}
\end{center}
\caption{a) A brane configuration for SU(5) theory with 7 fundamentals, here $k_{-}=2, k_{+}=5$; b) Another brane configuration for the same
theory as a), here $k_{-}=3, k_{+}=4$.  }
\end{figure}

Let's consider pure $SU(N)$ theory for an example. There are two irregular singularities with $n=N, k=0$ in
(\ref{higgs},\ref{higgs1}). The spectral curve around the singularity is described in
(\ref{pure}). We put two  singularities at $z=0, z=\infty$, and the Seiberg-Witten curve is
\begin{equation}
x^N+{u_2\over z^2}x^{N-2}+{u_3\over z^3}x^{N-3}+....+{u_{N-1}\over z^{N-1}}x+{\Lambda^N\over z^{N+1}}+{u_N\over z^N}+{\Lambda^N\over z^{N-1}}
=0.
\end{equation}
The Seiberg-Witten differential is $\lambda=xdz$.

The solution of pure $SU(N)$ theory is related to another integrable system: periodic Toda chain \cite{inte1,inte2,inte3}. Here we
find another integrable system to describe pure $SU(N)$ theory using Hitchin's system, these two integrable
systems should  be isomorphic.

The specific form of the singularity of the solution to Hitchin's equation may be seen
from the Brane configuration and Seiberg-Witten curve.
The Seiberg-Witten curve for the brane configuration in Figure 6 is
\begin{equation}
F(v,t)=c_0\prod_{i=1}^{k_{-}}(v-m_i)t^2+B(v)t+c_2\prod_{i=1}^{k_{+}}(v-m_i)=0, \label{sw}
\end{equation}
where $B(v)=c_1(v^N+u_2v^{N-2}+...u_N)$, and the Seiberg-Witten differential is $\lambda={v\over t}dt$.
We regard this curve as the polynomial in v with fixed t,
and there are a total of N roots. In the limit $t\rightarrow 0$, $k_{+}$ roots are constant, they
are
\begin{equation}
v_1=m_1,~v_2=m_2,...v_{k_+}=m_{k_{+}}.
\end{equation}

Since $k_{+}<N$, there are $(N-k_{+})$ roots which are not constant
\begin{equation}
\Lambda t^{1\over N-k_{+}}(1,\omega,...\omega^{N-k_{+}-1}),
\end{equation}
where $\omega$ is the root of the equation $x^{N-k_{+}}=1$, and $\Lambda$ depends on $c_\alpha$.

Now we want to find the Hitchin system description. The Seiberg-Witten curve is identified
with the spectral curve of Hitchin system
\begin{equation}
det(x-\Phi(z))=0,
\end{equation}
and the Seiberg-Witten differential is $\lambda=xdz$. Let's compare this with (\ref{sw}),
we identify $z$ with t, and $x={v\over t}$, the equation is
\begin{equation}
det({v\over t}-\Phi(t))\propto det(v-t\phi(t))=F(v,t).
\end{equation}
Now we can read the boundary condition of the Higgs field at the puncture $t=0$, The roots
of $v$ at fixed t are identified with the eigenvalues of the function $t\Phi(t)$. So the
Higgs field has the following form at $t=0$,
\begin{equation}
\Phi={1\over t^{1+{1\over N-k_{+}}}}diag(0,...0,\Lambda, \Lambda \omega,...\Lambda\omega^{N-k_{+}-1})dt+{1\over t}diag(m_1,m_2,..m_k,m_{k+1}, m_{k+1}...m_{k+1})dt+... \label{higgs2}.
\end{equation}

We do a little bit manipulation on the simple pole term so that the matrix is traceless, i.e. they are
taking value in the lie algebra of $SU(N)$. This is the exactly same as $(\ref{higgs}, \ref{higgs1})$. In the case $k_{+}=N$, the irregular term is absent and we have N mass terms, however, the maximal parameter for a regular singularity
is $N-1$, so we need to turn on another simple singularity with only one mass parameter. The analysis can be
carried out similarly for $t=\infty$.

We next turn to the description of the quiver gauge theory. The simplest case is a quiver with all $SU(N)$ gauge
groups, and only the two $SU(N)$ group at the ends are not conformal. If we have $n$ SU(N) gauge groups, then
there are $n-1$ simple punctures and two irregular punctures depending on the number of fundamentals on the end
$SU(N)$ gauge group. The dimension of the complex structure moduli space of this sphere with $n+1$  punctures
is $n-2$, this matches the number of conformal gauge group and is used to describe the UV conformal gauge
coupling constants. If there are $k_{-}$ fundamental hypermultiplets on the far left $SU(N)$ gauge group and $k_+$
fundamentals on the far right, the two irregular singularity is of the form $(\ref{higgs}, \ref{higgs1})$ with
$n=N, k=k_{\pm}$.

Let's consider the general quiver gauge theory with the gauge group
$\prod_{i=1}^n SU(k_i)$, here $k_1<k_2...<k_r=...k_s>k_{s+1}..>k_n$ and $k_s=..k_r=N$, the rank of the gauge group is chosen so that every gauge group is conformal or asymptotically free; there are bi-fundamental fields between adjacent gauge groups, we also add $d_i$ fundamental hypermultiplets to $i$th gauge group $SU(k_i)$. The $r-s-1$  middle $SU(N)$ gauge groups are conformal. The SCFT case is studied
in \cite{Gaiotto1,dan}, we want to extend it to general asymptotical free case. We review the superconformal
theory which will provide us a lot of clues, in this case $d_\alpha=2k_\alpha-k_{\alpha-1}-k_{\alpha+1}$.

The Seiberg-Witten curve for this theory is derived by lifting the brane configuration to M theory. The $D6$ branes
are described by Taub-NUT space \cite{witten3}. $NS5-D4$ brane configurations become a single M5 brane embedded in
D6 branes background. Define coordinate $v=x^4+ix^5$ and polynomials:
\begin{equation}
J_s=\prod_{a=i_{s-1}+1}^{i_s}(v-m_a),
\end{equation}
where $1\leq s \leq n$ and $d_\alpha=i_\alpha-i_{\alpha-1}$, $m_a$ is the constant
which represents the position of D6 brane in $v$ direction and is identified with
the mass of the fundamental hypermultiplet.

The Seiberg-Witten curve is
\begin{eqnarray}
t^{n+1}+g_1(v)t^n+g_2(v)J_1(v)t^{n-1}+g_3(v)J_1(v)^2J_2(v)t^{n-2}\nonumber\\
+...+g_\alpha\prod_{s=1}^{\alpha-1}J_s^{\alpha-s}t^{n+1-\alpha}+...+f\prod_{s=1}^nJ_s^{n+1-s}=0 \label{SW},
\end{eqnarray}
here $g_\alpha$ is a degree $k_\alpha$ polynomial of variable $v$. From the
study of a single $SU(N)$ theory, to get a Hitchin description, it is necessary to move all
the D6 branes to the far left and far right. We split $J_\alpha$ as the product of $J_{\alpha, L}$ and
$J_{\alpha, R}$, where $J_{\alpha, L}$ denotes the D6 branes moving to the left, and
$J_{\alpha, R}$ denotes the D6 branes moving to the right.
 $J_{\alpha, L}$ and $J_{\alpha, R}$ can be chosen arbitrarily. We define the canonical
choice by moving all the fundamental matters for $SU(k_i), i\leq r$ to the far left,
and move all the fundamental matters for $SU(k_i), i\geq s$ to the far right. In the case of
$r=s$, we split the fundamentals into two parts $d_{r,L}=N-k_{r-1}$ and $d_{r,R}=N-k_{r+1}$.

After moving the D6 branes to the infinity, the Seiberg-Witten curve becomes \cite{moore}
\begin{equation}
F(v,t)=\sum_{\alpha=0}^{n+1}\hat{g}_\alpha(v)t^{n+1-\alpha} \label{SW1}
\end{equation}
Where
\begin{equation}
\hat{g}_{\alpha}(v)=c_\alpha g_\alpha(v)\prod_{\beta=\alpha+1}^rJ_\beta^{\beta-\alpha}~~~~~~~~\alpha=0,1,....r-1
\end{equation}
\begin{equation}
\hat{g}(v)=c_\alpha g_\alpha(v),~~~~~~~~~~~~~~~~~~~~~~~~~~~~~~~~~~~~~~~~\alpha=r,....s
\end{equation}
\begin{equation}
\hat{g}(v)=c_\alpha g_\alpha(v) \prod_{\beta=s}^\alpha J_\beta^{\alpha-\beta}~~~~~~~~~~\alpha=s-1,.....n.
\end{equation}

Let's calculate the order of the polynomial $\hat{g}_\alpha (v)$. The middle one is not changed.
There are two quiver legs and we study one of them and the other leg can be treated similarly. Let's consider the leg $SU(k_1)-SU(k_2)-...-SU(k_r)$ with $k_r=N$, We can associate a Yang tableaux to this leg with the rows $n_1=k_1$, $n_2=k_2-k_1$,...,$n_r=k_r-k_{r-1}$, it is easy to see that the total box of the Yang tableaux is $N$. Then the number of fundamentals can be written as $d_\alpha=2k_\alpha-k_{\alpha-1}-k_{\alpha+1}=(k_\alpha-k_{\alpha-1})-(k_{\alpha+1}-k_{\alpha})=n_\alpha-n_{\alpha+1}$.
$J_\alpha(v)$ is a order $d_\alpha$ polynomial in $v$. The order of polynomial $\hat{g}_\alpha(v)$ is
\begin{equation}
d(g_\alpha)=k_\alpha+d_{\alpha+1}+2d_{\alpha+2}+...(r-\alpha)d_r=N
\end{equation}
The same calculation can be applied to the right quiver leg, so the polynomial $g_\alpha(v)$ has the same
order $N$. Let's study the roots of Seiberg-Witten curve regarded as a polynomial in $v$ with
$t$ fixed. $v$ have $N$ constant roots at the roots $t_\alpha$ of the following equation
\begin{equation}
\sum_{\alpha=0}^{n+1}c_\alpha t^{n+1-\alpha}=0.
\end{equation}
This polynomial is the coefficient of $v^N$ in $F(v,t)$ if we regard it as the polynomial in $v$.
The singular behavior of the Higgs field of the Hitchin equation can be derived from the roots as
we did for a single $SU(N)$ gauge group theory, the Higgs field has regular simple singularity.
The Higgs field has the form
\begin{equation}
\Phi(z)={1\over z}diag(\underbrace{m,...m}_{N-1},-(N-1)m)dz+...
\end{equation}

There are other singularities for the Hitchin system, we study the roots of the Seiberg-Witten curve
at $t\rightarrow \infty$, the $n$ roots of $v$ are dictated by the polynomial $g_0(v)$,
\begin{equation}
v=(\underbrace{m_{d_1,1},m_{d_1,2},..m_{d_1,d_1}}_{d_1}\underbrace{m_{d_2,1},m_{d_2,1},
...m_{d_2,d_2},m_{d_2,d_2}}_{2d_2},....).
\end{equation}
The Higgs field is therefore
\begin{equation}
\Phi(z)={1\over z}diag(\underbrace{m_{d_1,1},m_{d_1,2},..m_{d_1,d_1}}_{d_1}
\underbrace{m_{d_2,1},m_{d_2,1},...m_{d_2,d_2},m_{d_2,d_2}}_{2d_2},....)dz+...
\end{equation}
It is interesting to note that the mass pattern can be derived from the Young tableaux. Indeed,
it can be read from the dual Young tableaux. To construct dual Young tableaux, we simply
exchange the rows and columns. More precisely, the dual Young tableaux is related to the original Yang tableaux as follows: we have a total of $n_1$ rows, we have $n_1-n_2$ rows with length 1, $n_{k}-n_{k+1}$ rows with length k, etc; we use the convention $n_{r+1}=0$. Look at the form of the Higgs field, we see that the residue is dictated by
dual Young tableaux, we have $d_1=n_1-n_2$ mass parameters with degeneracy 1, we have $d_2=n_2-n_3$ mass
parameters with degeneracy 2, etc. The same analysis applies to the case $t\rightarrow 0$, so the theory is described
by a Riemann sphere with $n+1$ simple punctures and two generic punctures.

Let's now consider the non-conformal theory, here the number of fundamentals are
arbitrary and constrained by the relation $d_i\leq (2k_i-k_{i-1}-k_{i+1})$.
The Seiberg-Witten curve is the same as  $(\ref{SW1})$, the difference is that here $c_\alpha$ is dimensional parameters to make every term in $F(v,t)$ have the same dimension. Similarly, we have the simple singularity for the Higgs field at the points $t_\alpha$ which are the roots of the polynomial of the coefficient of the $v^N$ term. There
are other two singularities at $t=0$ and $t=\infty$, these two describe the right quiver tail and left quiver
tail respectively. We study the left quiver tail and the right quiver tail can be studied similarly. In the
case $r=s$, there is no canonical way to split the fundamental matters on the $SU(k_r$ node, but we have to make sure that $d_{rL}\leq N-k_{r-1}$, $d_{rR}\leq N-k_{r+1}$.

The order of coefficient for $\hat{g}_\alpha(v),~~\alpha<r$ is
\begin{equation}
dim(\hat{g}_\alpha(v))=k_\alpha+d_{\alpha+1}+2d_{\alpha+2}....(r-\alpha)d_r=\hat{k}_\alpha,
\end{equation}
$\hat{k}_\alpha$ is a non-decreasing series and
\begin{equation}
\hat{k}_{\alpha}-\hat{k}_{\alpha-1}=(k_\alpha-k_{\alpha-1})-\sum_{i=\alpha}^rd_i
=n_{\alpha}-\sum_{i=\alpha}^rd_i.
\end{equation}
We take $\hat{k}_{\alpha-1}=0$ and
we have the condition $\hat{k}_{\alpha}-\hat{k}_{\alpha-1}\geq 0$, notice that $\hat{k}_r=N$.

In the limit of $t\rightarrow \infty$, $v$ have $\hat{k}_0$ constant roots
\begin{equation}
v=(\underbrace{m_{d_1,1},m_{d_1,2},..m_{d_1,d_1}}_{d_1}\underbrace{m_{d_2,1},m_{d_2,1},
...m_{d_2,d_2},m_{d_2,d_2}}_{2d_2},....).
\end{equation}
Since $\hat{k}_0<N$, there are other roots besides the constant ones we get above, for any $\alpha\leq r$,
we have the $\hat{k}_\alpha-\hat{k}_{\alpha-1}$ roots
\begin{equation}
v=\Lambda_\alpha t^{1\over m}(1,\omega,...\omega^{m-1}),
\end{equation}
where $m=\hat{k}_\alpha-\hat{k}_{\alpha-1}$, $\omega$ is the root for $x^m=1$ and
$\Lambda_\alpha=({c_{\alpha-1}\over c_\alpha})^{1\over m}$, one can check that $\Lambda$ has
dimension 1 from the Seiberg-Witten curve $F(v,t)$, if we require $t$
has dimension $-1$ and $v$ has dimension $1$.

Based on the roots of $v$ in the limit $t\rightarrow \infty$, We propose that the Higgs field has the following form
around the singularity $z^{'}=\infty$ (we change the local coordinate to $z={1\over z^{'}}$
\begin{equation}
\phi(z)=\left(\begin{array}{ccccc}
v_{n_1}&0&0&0&0\\
0&v_{n_2}&0&0&0\\
0&0&...&0&0 \\
0&0&0&...&0\\
0&0&0&0&v_{n_r}\end{array} \right) \label{quiver}
\end{equation}
Where $v_{n_\alpha}$ is the diagonal matrix as in (\ref{higgs}) with $n=n_\alpha, k=\sum_{i=\alpha}^rd_i$, to make the Higgs field well defined, we make a gauge transformation of the block diagonal form
(\ref{mat}) when we cross the cut. The mass
terms in $v_{n_\alpha}$ is constrained though, its form is
\begin{equation}
{1\over z}diag(\underbrace{m_{d_\alpha,1},m_{d_\alpha,2},...,m_{d_\alpha,d_\alpha}}_{d_\alpha},...., \underbrace{m_{d_r,1},m_{d_r,2},...,m_{d_r,d_r}}_{d_r},m_\alpha,m_\alpha,...m_\alpha).
\end{equation}
It is interesting to compare the total mass parameters with the quiver tail. The mass parameters $(m_{d_1,1}..,m_{d_2,1},..,m_{d_r,1},...,m_{d_r,d_r})$ are used to describe the mass deformation for the fundamentals, and we have
a total of $r-1$ mass parameter $m_{\alpha}$ (One of $m_\alpha$ is eliminated by traceless condition), and
we have a total of $(r-1)$ bi-fundamental matter fields, so the mass parameters match the matter contents
of the quiver gauge theory.

If $n-k=1$ for $v_{n_\alpha}$, we need a little bit modification, the Higgs field is (we assume $n_r-k=1$ here for an illustration).
\begin{equation}
\Phi(z)=\left(\begin{array}{ccccc}
v_{n_1}+{1\over z^2}\Lambda_rI_{n_1\times n_1}&0&0&0&0\\
0&v_{n_2}+{1\over z^2}\Lambda_rI_{n_2\times n_2}&0&0&0\\
0&0&........&0&0 \\
0&0&0&v_{n_{r-1}}+{1\over z^2}\Lambda_rI_{n_{r-1}\times n_{r-1}}&0\\
0&0&0&0&{-(n-1)\over z^2}\Lambda_r\end{array} \right)dz+... \label{quiver}
\end{equation}
The coefficient $\Lambda_i$ is identified with the dynamical scale of $i$th gauge group $SU(k_i)$.

We need to clarify some of the special issues. There are at least $r-s-1$ simple regular
singularities which are used to describe the bi-fundamentals between the $SU(N)$ group. We may have more simple singularities if some of the gauge groups on the quiver tail is conformal.

If $n_{\alpha-1}-\sum_{i=\alpha-1}^rd_i\neq 0, n_{\alpha}-\sum_{i=\alpha}^rd_i=0$ for some $\alpha$, one can show that  all gauge groups $SU(k_i)$ with $r\geq i\geq \alpha$ are conformal, and only $\alpha-1$  blocks in the Higgs field are irregular, we need to add $(r-\alpha+1)$ more simple singularities to account for the mass deformation of
the bi-fundamental fields; this can also be seen from the fact that we have more roots for the equation
before $v^N$ term in $F(v,t)$. If $\alpha=1$ for the above situation, we return to the superconformal case. There is
another justification to add more simple singularities, since we only have $\alpha$ dynamical scale
from the irregular singularity, but we have extra $(r-\alpha+1)$ UV dimensionless
gauge couplings, these can only be represented by the complex structure moduli of the Riemann surface,
 so we need to add $(r-\alpha+1)$ simple regular punctures. We also lose mass parameters $m_\alpha$ for each regular block, these parameters are now encoded in the simple regular punctures.

What happens if there is a gauge group $SU(k_\beta)$ which is conformal, but $\beta<\alpha$, where for
$\alpha$,
\begin{equation}
n_{\alpha-1}-\sum_{i=\alpha-1}^rd_i\neq 0,~~~ n_{\alpha}-\sum_{i=\alpha}^rd_i=0.
\end{equation}
The above analysis implies that we need an irregular singularity and $(r-\alpha+1)$ simple regular singularities to
describe the quiver tail. We want to identify UV gauge couplings for $SU(k_\beta)$, it is not represented
by the complex structure moduli of the punctured Riemann sphere, it is encoded in the irregular part
of the irregular singularity.  The condition for the conformal gauge coupling of $SU(k_\beta)$ is $d_\beta=n_\beta-n_{\beta+1}$, the number of non-zero entries in irregular part of $v_{n_\beta}$ and
$v_{n_{\beta+1}}$ are
\begin{eqnarray}
r_\beta=n_\beta-\sum_{i=\beta}^rd_\beta,\nonumber\\
r_{\beta+1}=n_{\beta+1}-\sum_{i=\beta+1}^rd_\beta.
\end{eqnarray}
The conformal condition for $SU(k_\beta)$ implies $r_\beta=r_{\beta+1}$. Now the dimensionless
gauge coupling for $SU(k_\beta)$ is identified with $\tau_\beta={\Lambda_\beta\over\Lambda_{\beta+1}}={c_{\beta-1}c_{\beta+1}\over c_\beta^2}$. This
case shows that we can encode the conformal couplings in the irregular singularity, while in
the canonical treatment of superconformal field theory, the gauge coupling is encoded as
the complex structure moduli of the Riemann surface. Indeed, one can also encode all the
dimensional gauge couplings of the superconformal field theory into the irregular
singularity \cite{moore}.

The Seiberg-Witten curve has the familiar form
\begin{equation}
x^N=\phi_i(z)x^{N-i}.\label{gaiotto}
\end{equation}
The big difference with the conformal case is that in our case, not all the dimension $i$ Coulomb
branch parameters are encoded in the coefficient $\phi_i$. We would like to check that the base of
Hitchin's fibration matches the dimension of the Coulomb branch of the gauge theory. Let's do
this for the left quiver tail, in conformal case, it is described by 1 generic regular singularity
and $r$ simple regular singularities, the total Coulomb branch dimension from these punctures is
\begin{equation}
{1\over 2}[N^2-\sum_ir_i^2+r(2N-2)],
\end{equation}
where $r_i$ is the height of the $i$th column of Young tableaux. We would like to express it in terms
of the rows of the Young tableaux; This can be done by noting that we have $n_k-n_{k+1}$ columns with
height $k$, and the above formula becomes
\begin{equation}
{1\over 2}\sum_i n_i^2+\sum_{i< j}n_in_j-r+\sum_i(r-i+{1\over2})n_i. \label{dim}
\end{equation}

In the non-conformal cases, the same
dimension is described by a irregular singularity and several simple regular singularities. The spectral
curve around the irregular singularity is (we first assume that there is a total of $r$ irregular blocks
in irregular singularity).
\begin{equation}
\det(x-\phi(z))=\prod_i^r(x^{n_i}+{f(m_i)\over z}x^{n_i-1}+{u_2\over z^2}x^{n_i-2}+....{(\Lambda_i)^{n_i}\over z^{n_i+1}}+{u_n\over z^{n_i}}).
\end{equation}
Expanding the spectral curve as the form (\ref{gaiotto}), and find the maximal order of pole of $\phi_i$ which
do not solely depend on $\Lambda$ and $m$, we start with $N-n_1<j\leq N$, we choose the term
from $n_1$ factor and constant terms from other factors, the orders of pole of $\phi_j$ are given by
\begin{equation}
\sum_{i=2}^{r}n_i+r-1+1, \sum_{i=2}^{r-1}n_i+r-1+2,...,\sum_{i=2}^{r}n_i+r-1+n_r.
\end{equation}
The same analysis can also be carried out for $N-\sum_{i=1}^{k}n_i<j\leq  N-\sum_{i=1}^{k-1}n_i$ with $ 1\leq k\leq r$,
and $\phi_j$ has the following orders of pole
\begin{equation}
\sum_{i=k+1}^{r}n_i+r-k, \sum_{i=k+1}^{r}n_i+r-k+2,...,\sum_{i=k+1}^{r}n_i+r-k+n_k.
\end{equation}
So the total dimension is
\begin{equation}
\sum_{k=1}^{r}[\sum_{i=k}^{k-1}n_i+r-k+ \sum_{i=k+1}^{r}n_i+r-k+2+....+\sum_{i=k+1}^{r}n_i+r-k+n_k].
\end{equation}
After some calculation, the above expression becomes
\begin{equation}
{1\over 2}\sum_{k=1}^rn_k^2+\sum_{k<i}n_in_k-r+\sum_{k=1}^r(r-k+{1\over2})n_k.
\end{equation}
This is the same as $(\ref{dim})$. In the case when the irregular block for the irregular singularity is
less than $r$, we have extra simple singularities, one can also show that the
dimension of the sum of the irregular singularity and regular singularity matches
(\ref{dim}), see appendix I for details.

In summary, for $A$ type $N=2$ quiver gauge theory, the six dimensional description involves several regular
singularities and two irregular singularities. In the superconformal case, the irregular singularity
becomes also the regular singularity and they are of the special type which is dictated completely
by the rank of the gauge group. In the non-conformal cases, the irregular singularities
are determined by the rank of the gauge group and the number of the fundamentals. They can
be described uniformly.

Similarly, for the non-conformal quiver, one can study different degeneration limits and
study different dual frame of the same theory. In the completely degeneration limit, we can find
new theories without conventional lagrangian description.

\section{Conclusion}
In this paper, we discuss solution of Hitchin's equation used to describe
the general asymptotical free $N=2$ $A$ type quiver gauge theory. The Higgs field has
irregular singularity and we
give the explicit description of the singularity. In superconformal cases, the duality
of the gauge theory can be derived by studying different degeneration limits of the
Riemann surface on which we define the Hitchin equation, we can extend this analysis to
the asymptotical free case.

There are a lot of open questions deserving further research. For $SU(2)$ theory, there
are famous isolated Argyres-Douglas superconformal field theory. We conjecture that it
may be described by a $SU(2)$ Hitchin system with higher order singularity, It is interesting to
check whether this conjecture is true or not.

It is interesting to extend the study of matching Nekrasov partition function in gauge theory
with the conformal block in two dimensional conformal field theory. For $SU(2)$ cases, this
is studied by \cite{Gaiotto3}. Based on the characterization of the singularity, we can match the information
of the singularity to insertion of primary fields or irregular states. In the case of regular
singularity, this is studied in \cite{primary}. It is interesting to describe the irregular states based on
the information we describe for the irregular singularity.

It is shown in \cite{loop1} that Wilson loop and t'hooft loop of the $SU(2)$
quiver gauge theory can be classified
by studying the non-intersecting curve on punctured Riemann surface. We hope we can extend this
classification to asymptotical free case; The new feature is that there is
a cut around the irregular singularity. It is also interesting to calculate the expectation
value of the Wilson-t'hooft loop operators and the surface operator using conformal field
theories.

We can study four dimensional gauge theory on $\Omega$ deformation as in \cite{nek2,nek3,nek4} ,
and derive the twisted superpotential of effective two dimensional theory,
this will give the quantization of integrable system we found in this paper,
it is desirable to carry this calculation in detail.

Hitchin's moduli space has a hyperkahler metric, it is definitely giving us a lot of insights
if we can learn about the exact form of the metric (see the discussion in \cite{boalch1})
. The Hitchin's moduli space is the Coulomb
branch of three dimensional gauge theory derived by compactifying four dimensional gauge theory
on a circle. The metric of the moduli space plays a central role in proving the wall crossing
formula \cite{moore,moore1}. The irregular singularity we studied in this paper
is less well studied in Geometric Langlands Program \cite{wild}, it is interesting to
see whether those irregular singularities have special role in the context of
Geometric Langlands Program.

In this paper, we only study the canonical Hitchin system
representation of four dimensional $N=2$ gauge theory. There are more than one representations
for $SU(N)$ gauge theory with $N_f$ fundamentals even with canonical representation.
One would like to find an way to prove the isomorphism between those different looking
Hitchin moduli space. One way is to attach an graph \cite{boalch} to each Hitchin's
moduli space, if the graphs with different Hitchin system are same, we can conclude
they describe the same $N=2$ gauge theories. See another way of uniquely describing
$N=2$ gauge theory using surface operator \cite{loop4}.

It is interesting to extend the same analysis to four dimensional $A$ type quiver with $Usp$ and
$SO$ gauge group \cite{SO, SO1, SO2}. This involve six dimensional $D_N$ $(0,2)$ theory
compactified on a punctured Riemann surface. It is interesting to learn what type of regular
singularity and irregular singularity are needed to describe conventional quiver gauge theory
with $Usp$ and $SO$ group.

\begin{flushleft}
\textbf{Acknowledgments}
\end{flushleft}
It is a pleasure to thank F.Benini, D.Gaiotto, Y.Tachikawa for helpful discussions.
We thank Eric Mayes for carefully reading the draft.
This research was supported in part by the Mitchell-Heep chair in
High Energy Physics (CMC) and by the DOE grant DEFG03-95-Er-40917.

\begin{flushleft}
\textbf{Appendix I}
\end{flushleft}
In this appendix, we will check the dimension of the base to Hitchin's fibration matches
the Coulomb branch dimension of $A_N$ type quiver with gauge group $\prod_{i=1}^n SU(k_i)$
with $k_1<k_2<...<k_{r-1}<k_r=...=k_s>k_{s+1}>...>k_{n-1}>k_n$, here
$k_r=...k_s=N$. The matter contents are the bi-fundamental fields between
the adjacent gauge group and we add $d_i$ fundamental hypermultiplets on
each quiver node, $d_i$ is constrained so that the gauge theory is conformal
or asymptotical free.

As we discussed in the main part, the corresponding Hitchin system is defined
on a sphere with $r-s$ regular simple singularities to describe the bi-fundamental
fields between $SU(N)$ gauge group and two irregular singularities and several
simple regular singularities to describe two quiver tails.
For the left quiver tails $SU(k_1)-SU(k_2)-....-SU(k_r)$,
we need a irregular singularity and several regular simple singularities depending
on the number of fundamental hypermultiplets. The same analysis can be carried out
 for right quiver tail. For the conformal quiver, the left quiver tail is
described by $r$ simple singularities and 1 generic regular singularity, the contribution
to the dimension of the base of Hitchin's fibration from these singularities are
\begin{equation}
{1\over 2}\sum_i n_i^2+\sum_{i< j}n_in_j-r+\sum_i(r-i+{1\over2})n_i, \label{dim1}
\end{equation}
where $n_i=k_{i}-k_{i-1}$ is the partition of the Young tableaux associated with
the left quiver tail.

We want to check that in the non-conformal case with one irregular singularity
and several simple regular singularities, the dimensions from the local moduli
spaces are the same as (\ref{dim1}). We have confirmed this in the case with no
simple regular singularity, here we will check it for
the case with several simple singularities.

We first calculate the dimension from the irregular singularity. Let's define a
number for each quiver node:
\begin{equation}
p_i=n_i-\sum_{j=i}^rd_j .
\end{equation}
There is a number $\alpha$ such that $p_\alpha=0$ and $p_{\alpha-1}\neq 0$, and one can show that for any
$i>\alpha$, we have $p_i=0$. The Higgs field have the form
\begin{equation}
\Phi(z)=\left(\begin{array}{ccccc}
v_{n_1}&0&0&0&0\\
0&v_{n_2}&0&0&0\\
0&0&...&0&0 \\
0&0&0&...&0\\
0&0&0&0&v_{n_r}\end{array} \right)dz+... \label{quiver}
\end{equation}
Where $v_{n_i}$ is of the form $(\ref{higgs}, \ref{higgs1})$ with $n=n_i, k=\sum_{d=i}^r$,
in the case $i\geq \alpha$, we do not have the irregular part in $v_{n_i}$. The mass pattern of the
regular part is determined by the Young tableaux $Y^{'}$with partition $[n_\alpha,...,n_r]$.

The spectral curve around the irregular singularity has the decomposition
\begin{equation}
det(x-\phi(z)=\prod_{i=1}^{\alpha-1}(x^{n_i}+{f(m_i)\over z}x^{n_i-1}+{u_2\over z^2}x^{n_i-2}+....{(\Lambda_i)^{n_i}\over z^{n_i+1}}+{u_n\over z^{n_i}})g_{\alpha}
\end{equation}
Where $g_\alpha(x,z)$ is given by
\begin{equation}
g_\alpha(x,z)=x^{l}+{f(m)\over z}x^{l-1}+...({f_i(m)\over z^i}+{U_i\over z^{p_i}})x^{l-i}+...
\end{equation}
where $l=\sum_{i=\alpha}^r n_i$ and $p_i$ is determined by Young tableaux $Y^{'}$. The calculation for
the maximal poles for $ N-\sum_{i=1}^{\alpha-1}<j\leq N $ is similar to the one we did in the
main text, with the only difference that the constant term from $g_\alpha$ contributing an order
of $l$ instead of $l+r-\alpha$. For $N-\sum_{i=1}^{k}n_i<j\leq  N-\sum_{i=1}^{k-1}n_i$ with
$ 1\leq k\leq \alpha-1$, the orders of pole are
\begin{equation}
\sum_{i=k+1}^{r}n_i+\alpha-k-1, \sum_{i=k+1}^{r}n_i+\alpha-k-1+2,...,\sum_{i=k+1}^{r}n_i+\alpha-k-1+n_k.
\end{equation}
The total dimension for this range of $j$ is
\begin{equation}
{1\over 2}\sum_{k=1}^{\alpha-1}n_k^2+\sum_{i<k,~~i\leq \alpha-1}n_in_k+\sum_{k=1}^{\alpha-1}
(\alpha-1-k+{1\over2})n_k-(\alpha-1).
\end{equation}

For $j\leq N-\sum_{i=1}^{k}n_i$, the order of pole is given by the Young tableaux $Y^{'}$,
(this is similar to the regular pole case), the total dimension is
\begin{equation}
{1\over 2}\sum_{k=\alpha}^{r}n_k^2+\sum_{i<k,~~i> \alpha-1}n_in_k+\sum_{k=\alpha}^{r}
(\alpha-1-k+{1\over2})n_k.
\end{equation}

We also have $(r-\alpha+1)$ simple punctures with contribution to Coulomb branch
\begin{equation}
(r-\alpha+1)\sum_{k=1}^rn_k-(r-\alpha+1).
\end{equation}

Sum them up, the total dimension from the irregular singularity and
simple regular singularities are
\begin{equation}
{1\over 2}\sum_{k=1}^rn_k^2+\sum_{k<i}n_in_k-r+\sum_{k=1}^r(r-k+{1\over2})n_k,
\end{equation}
which is the same as ($\ref{dim1}$).

\end{document}